# R-factor analysis of data based on population models comprising R- and Q-factors leads to biased loading estimates

André Beauducel[*]

*Institute of Psychology, University of Bonn, Germany*

## Abstract

Effects of performing R-factor analysis of observed variables based on population models comprising R- and Q-factors were investigated. It was noted that estimating a model comprising R- and Q-factors has to face loading indeterminacy beyond rotational indeterminacy. Although R-factor analysis of data based on a population model comprising R- and Q-factors is nevertheless possible, this may lead to model error. Accordingly, even in the population, the resulting R-factor loadings are not necessarily close estimates of the original population R-factor loadings. It was shown in a simulation study that large Q-factor variance induces an increase of the variation of R-factor loading estimates beyond chance level. The results indicate that performing R-factor analysis with data based on a population model comprising R- and Q-factors may result in substantial loading bias. Tests of the multivariate kurtosis of observed variables are proposed as an indicator of possible Q-factor variance in observed variables as a prerequisite for R-factor analysis.

Keywords: R-factor analysis, Q-factor analysis, loading bias, model error, multivariate kurtosis

* Institute of Psychology, Rheinische Friedrich-Wilhelms-Universität Bonn, Kaiser-Karl-Ring 9, 53111 Bonn, Germany. Email: beauducel@uni-bonn.de

The factor model (Mulaik, 2012; Harman, 1976) allows for the investigation of measurement models in psychology and several areas of the social sciences. There are several estimation methods for the factor model, and researchers have the choice between several different methods for exploratory and confirmatory factor analysis (Flora & Curran, 2004; Mulaik, 2012; Muthén & Asparouhov, 2012; Wirth & Edwards, 2007). Although a very large number of studies is based on the factor model, the real-world phenomena may not correspond exactly to this model. Tucker and MacCallum (1991) emphasized that the factor model may not fit perfectly to real population data. The possible difference between the factor model and population real-world data has been termed 'model error' (Tucker & MacCallum, 1991; MacCallum, 2003). Accordingly, in a factor analysis performed on a real-world data sample, some misfit of the factor model might be due to sampling error and some misfit might be due to model error. The modeling of common and unique factors together with a large number of minor factors has successfully been used in order to generate more realistic data containing model error in simulation studies (e.g., de Winter, Dodou, & Wieringa, 2009). However, other types of model error that are not based on a large number of minor factors may also be relevant for the fit of the factor model to real data.

One form of model error considered here is that the covariances between observed variables may be affected by covariances between individuals. In psychology, factor analysis is mainly performed in order to identify latent variables explaining the covariation between variables that are observed in samples of individuals. Factor analysis of the covariances or correlations between variables that are observed across many individuals is often termed R-factor analysis whereas factor analysis of the covariances or correlations between individuals observed across many variables is termed Q-factor analysis (Ramlo & Newman, 2010; Broverman, 1961). A data matrix of individuals for Q-factor analysis is obtained when the matrix of observed variables used for R-factor analysis is transposed. Note that the empirical data used for R- and Q-factor analysis may be the same, although the number of observed variables will typically be larger than the number of individuals in Q-factor analysis whereas the number of individuals will typically be larger than the number of observed variables in R-factor analysis. Moreover, there are other preferences for factor extraction and rotation in Q-factor analysis (Akhtar-Danesh, 2016; Ramlo, 2016) than in R-factor analysis. Nevertheless, there is consensus that Q-factor analysis may be useful for the investigation of subjective individual views (Ramlo, 2016) and Q-factor analysis is sometimes preferred over R-factor analysis in the context of questionnaire development (e.g., Cadman, Belsky & Pasco Fearon, 2018).

The similarities and differences of R- and Q-factor analysis have primarily been discussed from the perspective of factor analysis as a tool for data analysis (Burt & Stephenson, 1939; Cattell, 1952). In consequence, the effects of the R- and Q-factor model as data generating population models on the results of R- or Q-factor analysis have rarely been in the focus of the investigation. It is therefore widely unknown what happens when data that are based on a population model comprising R- and Q-factors are submitted to R-factor analysis. As models are never true (Box, 1979) it is not the fact that model error occurs that is important here, but the question whether the loading estimates from R-factor analysis are substantially biased when a combined R- and Q-factor model holds. Therefore, and because most studies perform R-factor analysis, the focus of the present study is on the effect of a combined R- and Q-factor model as a population model on subsequent R-factor analysis. It is, however, acknowledged that a combined R- and Q-factor population model might also be a source of error for Q-factor analysis.

An example for R-factors in a context where Q-factors may also be relevant is the analysis of personality types in the context of personality traits (Gerlach, Farb, Revelle & Amaral, 2018), although the robustness of the results has been challenged (Freudenstein, Strauch, Mussel & Ziegler, 2019). Freudenstein et al. (2019) also noted that only 42% of the sample were associated with the proposed personality types indicating that the types are probably of moderate relevance. Although Gerlach et al. (2018) used cluster-methodology (Gaussian mixture models) for the identification of types, similarities of individuals have also been investigated by means of Q-factor analysis (Ramlo, & Newman, 2010). Thus, personality research shows that relevant similarities of variables as well as relevant similarities of individuals may co-occur. This does not imply that Q-factors yield a superior representation of personality variance nor that they allow for improved predictions of outcomes like, for example, social adjustment or job achievement (Asendorpf, 2003). For the present study it is only important to acknowledge that Q-factors may also be relevant for a complete description of the data. However, if we accept the idea that Q-factors may co-occur with R-factors, the consequences of a population model based on a combination of R- and Q-factors for the estimation of model parameters of R-factor analysis should be investigated. This has until now not been done as similarities of individuals have often been investigated by means of cluster analysis (Gerlach et al., 2018; Freudenstein et al., 2019), latent class analysis (Lazarsfeld & Henry, 1968), or factor mixture models (Lubke & Muthén, 2005). The achievements of these approaches for the analysis of typological variance are not questioned here. The focus of the present study is on the effect of population Q-factors co-occurring with population R-factors on the loading estimates of R-factor analysis which does not take into account the Q-factors.

After some definitions, the effects of population models based on R- and Q-factors on the covariance and correlation of observed variables and the resulting effects on the estimation of R-factor loadings are described for the population. Then, a simulation study is performed in order to give an account of the effect of population models comprising R- and Q-factors on loading estimates of R-factor analysis. Finally, a method indicating whether a data set contains a relevant amount of Q-factor variance is proposed and demonstrated by means of simulated data sets.

## Definitions

Let $\mathbf{X}_R$ be a $p \times n$ matrix of $p$ variables observed for $n$ individuals (Harman, 1976). The R-factor model can then be written as

$$\mathbf{X}_R = \mathbf{\Lambda}_R \mathbf{f}_R + \mathbf{\Psi}_R \mathbf{e}_R, \tag{1}$$

where $\mathbf{f}_R$ is a $q_R \times n$ matrix of normally distributed common R-factor scores, $\mathbf{\Lambda}_R$ is a $p \times q_R$ matrix of common R-factor loadings, $\mathbf{e}_R$ a $p \times n$ matrix of normally distributed linear independent unique R-factor scores, and $\mathbf{\Psi}_R$ is a $p \times p$ diagonal positive definite matrix of unique R-factor loadings. It is furthermore assumed that $E(\mathbf{f}_R) = \mathbf{0}$, $E(\mathbf{f}_R \mathbf{f}_R^{'}) = \mathbf{I}_{q_R}$, $E(\mathbf{e}_R) = \mathbf{0}$, $E(\mathbf{f}_R \mathbf{e}_R^{'}) = \mathbf{0}$, and $E(\mathbf{e}_R \mathbf{e}_R^{'}) = \mathbf{I}_p$, so that

$$\mathbf{\Sigma}_R = E(\mathbf{X}_R \mathbf{X}_R^{'}) = \mathbf{\Lambda}_R \mathbf{\Lambda}_R^{'} + \mathbf{\Psi}_R^2. \tag{2}$$

Let $\mathbf{X}_Q$ be a $n \times p$ matrix of $n$ individuals for which $p$ variables were observed. The Q-factor model can then be written as

$$\mathbf{X}_Q = \mathbf{\Lambda}_Q \mathbf{f}_Q + \mathbf{\Psi}_Q \mathbf{e}_Q, \tag{3}$$

where $\mathbf{f}_Q$ is a $q_Q \times p$ matrix of normally distributed common Q-factor scores, $\mathbf{\Lambda}_Q$ is a $n \times q_Q$ matrix of common Q-factor loadings, $\mathbf{e}_Q$ is a $n \times p$ matrix of normally distributed linear independent unique Q-factor scores, and $\mathbf{\Psi}_Q$ is a $n \times n$ diagonal positive definite matrix of unique Q-factor loadings. It is furthermore assumed that $E(\mathbf{f}_Q) = 0$, $E(\mathbf{f}_Q \mathbf{f}_Q^{'}) = \mathbf{I}_{q_Q}$, $E(\mathbf{e}_Q) = 0$, $E(\mathbf{f}_Q \mathbf{e}_Q^{'}) = \mathbf{0}$, $E(\mathbf{e}_Q \mathbf{e}_Q^{'}) = \mathbf{I}_n$, so that

$$\boldsymbol{\Sigma}_Q = E(\mathbf{X}_Q \mathbf{X}_Q^{'}) = \boldsymbol{\Lambda}_Q \boldsymbol{\Lambda}_Q^{'} + \boldsymbol{\Psi}_Q^2. \tag{4}$$

It is assumed that the observed variables $\mathbf{X}_R$ and $\mathbf{X}_Q$ are statistically independent with

$$
\begin{aligned}
E(\mathbf{X}_R \mathbf{X}^{'}_Q) &= E(\boldsymbol{\Lambda}_R \mathbf{f}_R + \boldsymbol{\Psi}_R \mathbf{e}_R)(\mathbf{f}^{'}_Q \boldsymbol{\Lambda}^{'}_Q + \mathbf{e}^{'}_Q \boldsymbol{\Psi}_Q))(p-1)^{-1} \\
&= E(\boldsymbol{\Lambda}_R \mathbf{f}_R \mathbf{f}^{'}_Q \boldsymbol{\Lambda}^{'}_Q + \boldsymbol{\Lambda}_R \mathbf{f}_R \mathbf{e}^{'}_Q \boldsymbol{\Psi}_Q + \boldsymbol{\Psi}_R \mathbf{e}_R \mathbf{f}^{'}_Q \boldsymbol{\Lambda}^{'}_Q + \boldsymbol{\Psi}_R \mathbf{e}_R \mathbf{e}^{'}_Q \boldsymbol{\Psi}_Q)(p-1)^{-1} = \mathbf{0},
\end{aligned}
\tag{5}
$$

for $E(\mathbf{X}^{'}_Q) = \mathbf{0}$ and with $E(\mathbf{f}_R \mathbf{f}^{'}_Q) = \mathbf{0}$, $E(\mathbf{f}_R \mathbf{e}^{'}_Q) = \mathbf{0}$, $E(\mathbf{e}_R \mathbf{f}^{'}_Q) = \mathbf{0}$, and $E(\mathbf{e}_R \mathbf{e}^{'}_Q) = \mathbf{0}$.

## A combined model of R- and Q-factors

The data in the following section are assumed to be analyzed from the perspective of R-factor analysis whereas the observed variables $\mathbf{X}_{RQ}$ are based on an aggregation of variables resulting from R- and Q-factors. This can be written as

$$\mathbf{X}_{RQ} = \mathbf{X}_R + \mathbf{X}^{'}_Q \mathbf{C}_n, \tag{6}$$

where $\mathbf{X}_R$ represents the part of the observed variables based on R-factors and $\mathbf{X}^{'}_Q$ is the transposed matrix of observed individuals based on Q-factors. Although adding $\mathbf{X}_R$ and $\mathbf{X}^{'}_Q$ is only possible for $n = p$, it should be noted that -in the combined model of R- and Q-factors- only $\mathbf{X}_{RQ}$ is observed whereas $\mathbf{X}_R$ and $\mathbf{X}^{'}_Q$ are parts of the assumed population model. Therefore, not all R- and Q-factors need not to be well represented by the observed variables in $\mathbf{X}_{RQ}$ when R-factor analysis is performed. For a complete description of the population model $n = p$ is nevertheless assumed in the following. Moreover, as $E(\mathbf{X}^{'}_Q)$ is not necessarily zero, there is the symmetric and idempotent centering matrix $\mathbf{C}_n = \mathbf{I}_n - n^{-1}\mathbf{1}_n \mathbf{1}^{'}_n$, based on the $n \times n$ identity matrix $\mathbf{I}_n$ and the $n \times 1$ column unit-vector $\mathbf{1}_n$, for row mean centering of $\mathbf{X}^{'}_Q$ on the right side of Equation 6. It has been noted by Cattell (1952) and others that mean centering of $\mathbf{X}^{'}_Q$ implies that the variance that would be based on a single common factor ($q_Q = 1$) in $\mathbf{X}_Q$ would be eliminated in R-factor analysis of $\mathbf{X}_{RQ}$. Therefore, only the condition $q_Q > 1$ is considered here.

It follows from Equation 6 that the covariances of $\mathbf{X}_{RQ}$ are

$$\mathbf{\Sigma}_{RQ} = E(\mathbf{X}_{RQ}\mathbf{X}'_{RQ}) = E(\mathbf{X}_R\mathbf{X}'_R + \mathbf{H}_{RQ} + \mathbf{H}'_{RQ} + \mathbf{X}'_Q\mathbf{C}_n\mathbf{X}_Q), \qquad (7)$$

with $\mathbf{H}_{RQ} = \mathbf{X}_R\mathbf{X}_{CQ}$ and $\mathbf{X}_{CQ} = \mathbf{C}_n\mathbf{X}_Q$. The element in the first row and first column of $\mathbf{H}_{RQ}$ is computed as $h_{RQ11} = \{x_{R11}x_{CQ11} + x_{R12}x_{CQ21} + \ldots + x_{R1n}x_{CQn1}\}$. As $\mathbf{X}_R$ and $\mathbf{X}_{CQ}$ are mean-centered, symmetrically distributed and as $E(\mathbf{X}_R\mathbf{X}'_Q) = \mathbf{0}$ all elements in the brackets are from the normal product distribution (Aja-Fernández, S. & Vegas-Sánchez-Ferrero, 2016; Weisstein, 2021), which is symmetric so that $E(h_{RQ11}) = 0$. This holds for all elements of $\mathbf{H}_{RQ}$ so that $E(\mathbf{H}_{RQ}) = \mathbf{0}$. Therefore, Equation 7 can be written as

$$\mathbf{\Sigma}_{RQ} = E(\mathbf{\Lambda}_R\mathbf{\Lambda}'_R + \mathbf{\Psi}^2_R) + E(\mathbf{f}'_Q\mathbf{\Lambda}'_Q\mathbf{C}_n\mathbf{\Lambda}_Q\mathbf{f}_Q + \mathbf{e}'_Q\mathbf{\Psi}_Q\mathbf{C}_n\mathbf{\Psi}_Q\mathbf{e}_Q). \qquad (8)$$

Two implications of assuming this model based on generating R- and Q-factors will be outlined here. The first consequence is the indeterminacy of the loadings resulting from R-factor analysis, the second consequence is the difference between the population R-factor loadings and the loading estimates resulting from R-factor analysis, which will be denoted as bias in the following.

*Indeterminacy of estimated R-factor loadings*

It follows from Equation 8 that the non-diagonal elements of $\mathbf{\Sigma}_{RQ}$ do not only depend on $\mathbf{\Lambda}_R$ but also on $\mathbf{\Lambda}_Q, \mathbf{\Psi}_Q, \mathbf{f}_Q$, and $\mathbf{e}_Q$. When R-factor analysis is performed for $\mathbf{\Sigma}_{RQ}$, the resulting estimated population loading estimates $\tilde{\mathbf{\Lambda}}_R$ will therefore not only have rotational indeterminacy but will also depend on the size of the elements in $\mathbf{\Psi}_R, \mathbf{\Lambda}_Q, \mathbf{\Psi}_Q, \mathbf{f}_Q$, and $\mathbf{e}_Q$, which typically remain unknown in empirical settings. The number of model parameters is much larger than the number of elements of $\mathbf{\Sigma}_{RQ}$. As the variances and covariances of the observed variables are relevant, only the elements of the lower triangle and the main diagonal of $\mathbf{\Sigma}_{RQ}$ represent independent data points, so that there are $(p^2 + p)/2$ independent data points. Although the number of parameters of the R-factor model (Equation 1) is $pq_R + p$, the parameters are typically identified successively, so that $pq_R$ common factor loadings are estimated first, and the $p$ unique factor loadings are calculated from the common factor loadings. As $q_R$ is typically considerably smaller than $p$ the factor model can easily be estimated. However, the number of elements of the model comprising R- and Q-factors (Equation 8) can be calculated from the number of elements in the matrices $\mathbf{\Lambda}_R$, $\mathbf{\Psi}_R$, $\mathbf{\Lambda}_Q$, $\mathbf{\Psi}_Q$, $\mathbf{f}_Q$, and $\mathbf{e}_Q$ is $pq_R + p + nq_Q + n + pq_Q + pn$. Even when $\mathbf{\Psi}_R$ and $\mathbf{\Psi}_Q$, the loadings of the unique factors are calculated in a second step, there are

$pq_R + nq_Q + pq_Q + pn$ model parameters. In order to give a further account of the number of

model parameters it is helpful to consider the condition of $n = p$, $q_R = q_Q$, this condition yields

$pq_R + pq_R + pq_R + p^2 = p^2 + 3pq_R$ model parameters. It follows that even for $q_R = 1$ the number

of model parameters is $p^2 + 3p$ which is more than two times $(p^2 + p)/2$, the number of data points. Thus, even the simplest combined R- and Q-factor model has considerably more parameters than there are elements in $\mathbf{\Sigma}_{RQ}$. Therefore, the combined R- and Q-factor model is

inherently non-identified and indeterminate, unless further constraints are imposed. Performing

R-factor analysis is nevertheless possible because the loadings and scores of the generating Q-

factors are simply ignored. However, as will be investigated in the following, the incomplete estimation of the parameters of the generating R- and Q-factor model by means of R-factor analysis may result in biased model parameters.

*Bias of estimated R-factor loadings*

For $\mathbf{X}_Q^{'}$ being mean centered ( $E(\mathbf{X}_Q^{'}) = 0$ ) Equation 8 implies that the variance of the elements in $\mathbf{\Sigma}_{RQ}$ is also affected by Q-factors. For $E(\mathbf{X}_Q^{'}) = 0$ the numerator of the variance of the

elements in $\mathbf{e}_Q^{'}\mathbf{\Psi}_Q^2\mathbf{e}_Q$ is

$$SSQ(\mathbf{e}_Q^{'}\mathbf{\Psi}_Q^2\mathbf{e}_Q) = tr((\mathbf{e}_Q^{'}\mathbf{\Psi}_Q^2\mathbf{e}_Q - E(\mathbf{e}_Q^{'}\mathbf{\Psi}_Q^2\mathbf{e}_Q))(\mathbf{e}_Q^{'}\mathbf{\Psi}_Q^2\mathbf{e}_Q - E(\mathbf{e}_Q^{'}\mathbf{\Psi}_Q^2\mathbf{e}_Q))^{'}), \qquad (9)$$

where SSQ denotes the sum of squares. It follows from $E(\mathbf{X}_Q^{'})=\mathbf{0}$, $E(\mathbf{e}_Q^{'}\mathbf{e}_Q)=\mathbf{0}$, $E(\mathbf{e}_Q^{'}\mathbf{\Psi}_Q^2\mathbf{e}_Q)=\mathbf{0}$, and $E(\mathbf{e}_Q\mathbf{e}_Q^{'})=\mathbf{I}_n$ that the eigen-decomposition of $\mathbf{e}_Q^{'}\mathbf{\Psi}_Q^2\mathbf{e}_Q=\mathbf{K}_Q\mathbf{V}_Q\mathbf{K}_Q^{'}$, where $\mathbf{V}_Q$ is a $n \times n$ identity matrix containing the eigenvalues in the main diagonal in descending order with $\mathbf{K}_Q\mathbf{K}_Q^{'}=\mathbf{K}_Q^{'}\mathbf{K}_Q=\mathbf{I}_n=\mathbf{V}_Q$. According to Magnus and Neudecker (2007, p. 248) the trace of the power of a positive semidefinite square matrix is equal to the trace of the power of the eigenvalues of the matrix so that

$$SSQ(\mathbf{e}_Q^{'}\mathbf{\Psi}_Q^2\mathbf{e}_Q) = tr(\mathbf{e}_Q^{'}\mathbf{\Psi}_Q^2\mathbf{e}_Q\mathbf{e}_Q^{'}\mathbf{\Psi}_Q^2\mathbf{e}_Q) = tr(\mathbf{V}_Q^2) = tr(\mathbf{\Psi}_Q^2). \tag{10}$$

When all unique Q-factors and all common Q-factors account for the same amount of variance of each observed variable ($1/2\, diag(\mathbf{\Sigma}_Q) = diag(\mathbf{\Lambda}_Q\mathbf{\Lambda}_Q^{'}) = \mathbf{\Psi}_Q^2$), the right-hand side of Equation 10 can be written as

$$tr(\mathbf{\Psi}_Q^2) = \tfrac{1}{2}n\sigma_Q^2 \tag{11}$$

It follows from $SSQ(\mathbf{e}_Q^{'}\mathbf{\Psi}_Q^2\mathbf{e}_Q) > 0$ that $\mathbf{e}_Q^{'}\mathbf{\Psi}_Q^2\mathbf{e}_Q$ introduces variability into the elements of $\mathbf{\Sigma}_{RQ}$

For $E(\mathbf{X}_Q^{'}) = 0$ the numerator of the variance of $\mathbf{f}_Q^{'}\mathbf{\Lambda}_Q^{'}\mathbf{\Lambda}_Q\mathbf{f}_Q$ is

$$SSQ(\mathbf{f}_Q^{'}\mathbf{\Lambda}_Q^{'}\mathbf{\Lambda}_Q\mathbf{f}_Q) = tr((\mathbf{f}_Q^{'}\mathbf{\Lambda}_Q^{'}\mathbf{\Lambda}_Q\mathbf{f}_Q - E(\mathbf{f}_Q^{'}\mathbf{\Lambda}_Q^{'}\mathbf{\Lambda}_Q\mathbf{f}_Q))(\mathbf{f}_Q^{'}\mathbf{\Lambda}_Q^{'}\mathbf{\Lambda}_Q\mathbf{f}_Q - E(\mathbf{f}_Q^{'}\mathbf{\Lambda}_Q^{'}\mathbf{\Lambda}_Q\mathbf{f}_Q))^{'}). \tag{12}$$

It follows from $E(\mathbf{f}_Q^{'}\mathbf{\Lambda}_Q^{'}\mathbf{\Lambda}_Q\mathbf{f}_Q)=\mathbf{0}$ that the eigen-decomposition of $\mathbf{f}_Q^{'}\mathbf{\Lambda}_Q^{'}\mathbf{\Lambda}_Q\mathbf{f}_Q=\mathbf{L}_Q\mathbf{W}_Q\mathbf{L}_Q^{'}$, where $\mathbf{W}_Q$ is a $n \times n$ diagonal matrix with $q_Q$ non-zero eigenvalues in decreasing order and $\mathbf{L}_Q\mathbf{L}_Q^{'}=\mathbf{L}_Q^{'}\mathbf{L}_Q=\mathbf{I}_{q_Q}$. The numerator of the variance of the elements of $\mathbf{f}_Q^{'}\mathbf{\Lambda}_Q^{'}\mathbf{\Lambda}_Q\mathbf{f}_Q$ is

$$SSQ(\mathbf{f}_Q^{'}\mathbf{\Lambda}_Q^{'}\mathbf{\Lambda}_Q\mathbf{f}_Q) = tr(\mathbf{f}_Q^{'}\mathbf{\Lambda}_Q^{'}\mathbf{\Lambda}_Q\mathbf{f}_Q\mathbf{f}_Q^{'}\mathbf{\Lambda}_Q^{'}\mathbf{\Lambda}_Q\mathbf{f}_Q) = tr(\mathbf{W}_Q^2). \tag{13}$$

which implies that the variance of the elements in $\mathbf{f}_Q^{'}\mathbf{\Lambda}_Q^{'}\mathbf{\Lambda}_Q\mathbf{f}_Q$ is greater zero. When all unique Q-factors and all common Q-factors account for the same amount of variance of each observed variable ($1/2\, diag(\mathbf{\Sigma}_Q) = diag(\mathbf{\Lambda}_Q\mathbf{\Lambda}_Q^{'}) = \mathbf{\Psi}_Q^2$), the right-hand side of Equation 13 can be written as

$$tr(\mathbf{W}_Q^2) = tr(\frac{n^2}{2q_Q^2}\sigma_Q^2) = \frac{n^2}{2q_Q}\sigma_Q^2. \tag{14}$$

It follows from Equations 14 and 10 and for $n > q_Q$ that $\sigma_Q^2 n^2 \left(2 q_Q\right)^{-1} > 0.5 n \sigma_Q^2$, i.e., that common Q-factors introduce $n/q_Q$ times more variability into the elements of $\boldsymbol{\Sigma}_{RQ}$ than unique Q-factors. More generally, Equations 10 and 11 imply that some variability in the elements of $\boldsymbol{\Sigma}_{RQ}$ is introduced by the common and unique Q-factors.

To sum up, Q-factors tend to enhance the variance of the covariances of observed variables (Equations 10, 11). However, the abovementioned analyses do not inform on the size of the respective effects and which amount of Q-factor variance might substantially distort an R-factor solution.

## Simulation study on the effect of generating Q-factors on R-factor loadings

A simulation was performed in order to give an account of the bias of R-factor loadings that is due to Q-factors. As the number of individuals or cases $n$ is part of the Q-factor model, the population has to comprise a large number of samples of a given $n$. The first population was based on 2,000 samples of $n = 300$ cases, the second population comprised 2,000 samples of $n = 600$ cases, and the third population comprised 2,000 samples of $n = 900$ cases. Accordingly, the conditions of the simulation study were $q_R = 3$, $q_Q = 3$, and $p = 15$. To investigate the effect of Q-factors on the variability of R-factor loading estimates, the salient loading sizes were set equal within each population model. The size of salient loadings in the common R-factor loading matrices $\boldsymbol{\Lambda}_R$ was $\lambda_R \in \{.50, .70\}$ and the size of salient loadings in common Q-factor loading matrices $\boldsymbol{\Lambda}_Q$ was $\lambda_Q = .90$. The non-salient loadings were zero in all population models. According to Equations 1, 3 and 6, the R- and Q-factor loadings were combined in order to generate the observed variables. This can be written as

$$\mathbf{X}_{RQ} = \boldsymbol{\Lambda}_R \mathbf{f}_R + \boldsymbol{\Psi}_R \mathbf{e}_R + (\mathbf{f}_Q' \boldsymbol{\Lambda}_Q' + \mathbf{e}_Q' \boldsymbol{\Psi}_Q) \mathbf{C}_n. \tag{15}$$

Although the relative effect of R- and Q-factors can be determined by the size of the respective common and unique R- and Q-factor loadings, it is helpful to control for the relative effect of R- and Q-factors more directly by means of

$$\mathbf{X}_{RQ} = \mathbf{\Lambda}_R \mathbf{f}_R + \mathbf{\Psi}_R (w_R \mathbf{e}_R + w_Q \mathbf{D}^{-0.5} (\mathbf{f}_Q' \mathbf{\Lambda}_Q' + \mathbf{e}_Q' \mathbf{\Psi}_Q) \mathbf{C}_n), \tag{16}$$

with $1 = w_R^2 + w_Q^2$ and $\mathbf{D} = diag((\mathbf{f}_Q' \mathbf{\Lambda}_Q' + \mathbf{e}_Q' \mathbf{\Psi}_Q) \mathbf{C}_n \mathbf{C}_n' (\mathbf{\Lambda}_Q \mathbf{f}_Q + \mathbf{\Psi}_Q \mathbf{e}_Q))$, which is needed to standardize the transposed part of the observed variables based on Q-factors. The usual metric of standardized factor loadings was maintained in the population with $\mathbf{I}_p = diag(\mathbf{\Lambda}_R \mathbf{\Lambda}_R' + \mathbf{\Psi}_R^2)$ and $\mathbf{I}_n = diag(\mathbf{\Lambda}_Q \mathbf{\Lambda}_Q' + \mathbf{\Psi}_Q^2)$. The observed variables were computed from Equation 16 by means of $q_R$ common factor scores $\mathbf{f}_R$, $p$ unique factor scores $\mathbf{e}_R$, $n/q_Q$ common factor scores $\mathbf{f}_Q$, and $n$ unique factor scores $\mathbf{e}_Q$, which were generated from normal distributions with $\mu = 0$ and $\sigma = 1$ by the Mersenne twister random number generator integrated in IBM SPSS, Version 26.0.

For $w_R^2 = 1.00$ and $w_Q^2 = .00$ Equation 16 yields a conventional R-factor model. For $w_R^2 = .50$ and $w_Q^2 = .50$, half of the unique R-factor variance is replaced by common and unique Q-factor variance. Four levels of $w_R^2$ (1.00, .75, .50, and .25) with the corresponding $w_Q^2$ were combined with two levels of $\lambda_R$ and three sample sizes $n$, which leads to $4 \times 2 \times 3 = 24$ populations, each comprising 2,000 samples.

Each set of $p$ observed variables was submitted to R-factor analysis. The dependent variables of the simulation study were the mean and standard deviation of the estimated loadings $\hat{\mathbf{\Lambda}}_R$ resulting from principal-axis R-factor analysis of the sample data with subsequent orthogonal target-rotation (Schönemann, 1966) of the estimated R-factor loadings $\hat{\mathbf{\Lambda}}_R$ towards the R-factor loadings $\mathbf{\Lambda}_R$ of the population model based on R- and Q-factors. Therefore, differences between the means of $\hat{\mathbf{\Lambda}}_R$ and cannot be due to different rotations of the factors.

*Results*

The most important result of the simulation study is that the standard deviation of the salient loadings increases with decreasing $w_R^2$ (Table 1). The results of $w_R^2 = 1.00$ show the standard deviations of the loadings that are only due to sampling error, as rotational variation of loadings was excluded by means of orthogonal target-rotation towards the population loadings. Especially, the results of $w_R^2 = 0.25$ show that the standard deviation of the loading estimates was about twice as large as the variation due to sampling error, when there was a substantial amount of Q-factor variance. This additional loading variation is a bias of the loading estimates as there was no salient loading variation in the population.

**Table 1.** Mean and standard deviation of target-rotated salient loading estimates of R-factor analysis for $\lambda_R = .50$, $.70$ and $\lambda_Q = .90$ for $w_R^2 = 0.25, 0.50, 0.75,$ and $1.00$ ($n = 600, 900$)

|           | $\lambda_R = .50$ |           |           | $\lambda_R = .70$ |           |           |
|-----------|-----------|-----------|-----------|-----------|-----------|-----------|
| $w_R^2$   | 300       | 600       | 900       | 300       | 600       | 900       |
| 0.25      | .50 / .12 | .50 / .08 | .50 / .07 | .70 / .06 | .70 / .04 | .70 / .03 |
| 0.50      | .50 / .09 | .50 / .06 | .50 / .05 | .70 / .05 | .70 / .03 | .70 / .03 |
| 0.75      | .50 / .07 | .50 / .05 | .50 / .04 | .70 / .04 | .70 / .03 | .70 / .02 |
| 1.00      | .50 / .06 | .50 / .04 | .50 / .04 | .70 / .04 | .70 / .02 | .70 / .02 |

*Note.* Standard deviations are given behind the slash.

In order to show the possible effect of the loading variation (comprising salient and non-salient loadings) on possible factor identification, a scatterplot of the target-rotated loadings of factor 1 and 2 is presented for $\lambda_R = .50$ in Figure 1 and for $\lambda_R = .70$ in Figure 2. Obviously, the overlap of salient and non-salient loadings for samples of $n = 300$ cases is substantial for $w_R^2 = 0.25$ and might be an obstacle for factor identification. In contrast, salient and non-salient loadings can clearly be separated for $n = 300$ cases, $\lambda_R = .50$ and $w_R^2 = 1.00$ or for samples sizes of $n = 600$ and $n = 900$. For all conditions based on $\lambda_R = .70$, the overlap of salient and non-salient loadings was small, indicating that factor identification would be possible (Figure 2). To sum up, when a substantial amount of Q-factor variance is expected, large sample sizes should be analyzed or very large R-factor loadings should be the expected as a basis for successful factor identification.

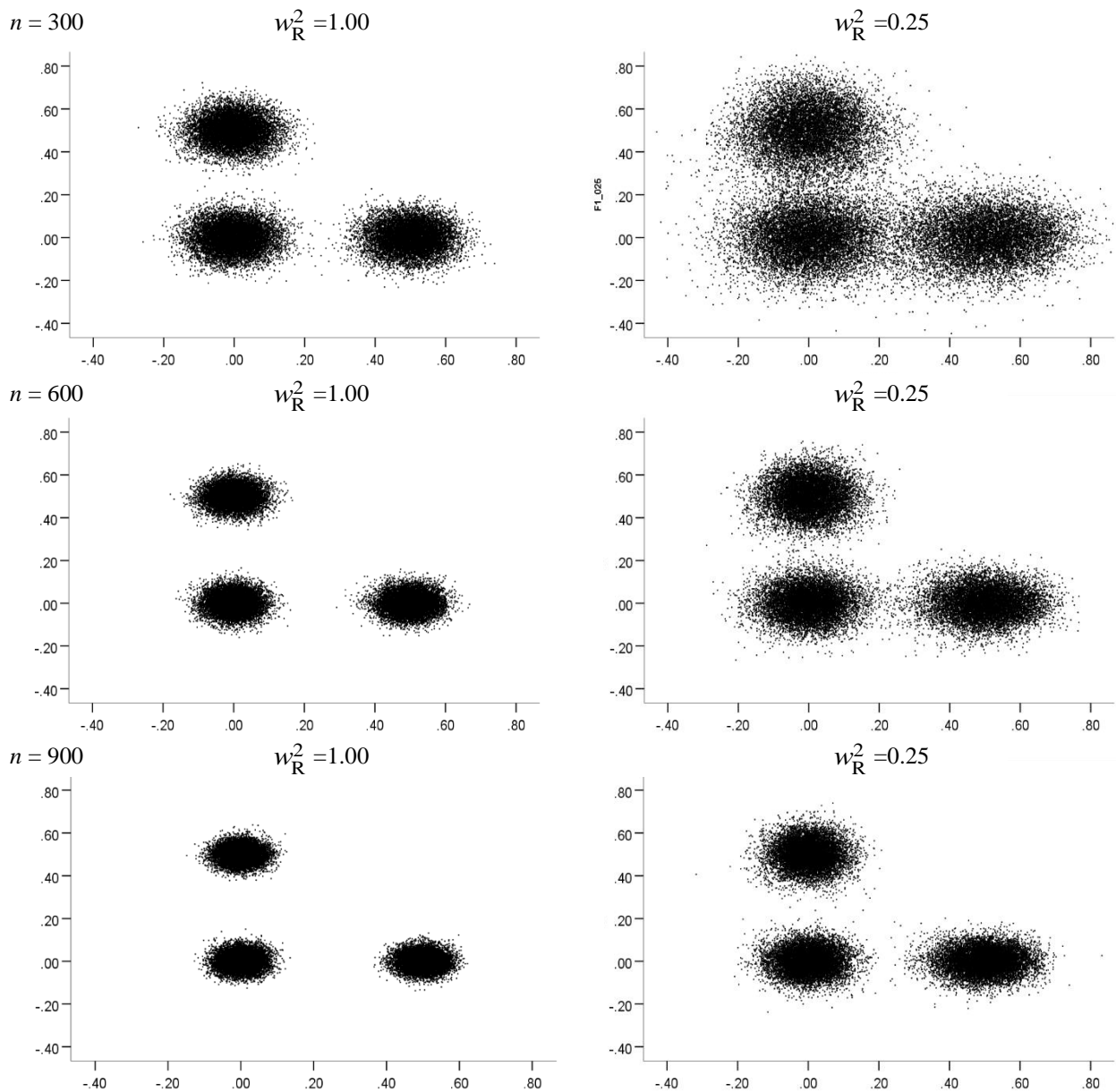

**Figure 1**. Scatterplot of R-factor loading estimates of factor 1 and 2 based on 2,000 samples ($n = 600$, $n = 900$) drawn from populations based on $\lambda_R = .50$, $q_R = 3$ R-factors ( $w_R^2 = 1.00$) and from populations comprising $q_R = 3$ R- and $q_Q = 3$ Q-factors ( $w_R^2 = 0.25$)

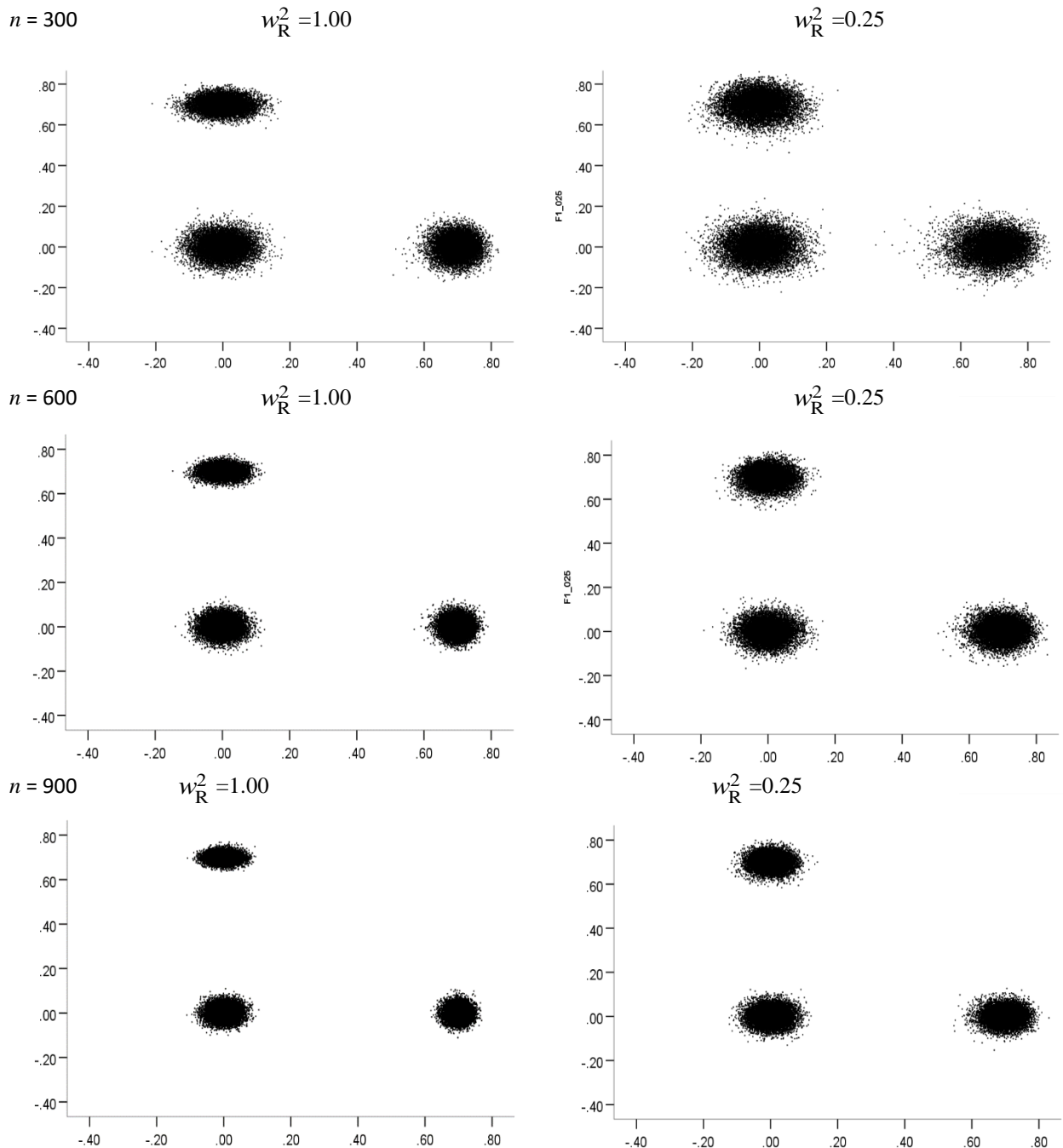

**Figure 2**. Scatterplot of R-factor loading estimates of factor 1 and 2 based on 2,000 samples ($n = 600$, $n = 900$) drawn from populations based on $\lambda_R = .70$, $q_R = 3$ R-factors ($w_R^2 = 1.00$) and from populations comprising $q_R = 3$ R- and $q_Q = 3$ Q-factors ($w_R^2 = 0.25$)

### An indicator of Q-factor variance

As R-factor analysis of data from a population based on a relevant amount of Q-factor variance may result in biased R-factor loadings, it is interesting to know whether there is a relevant amount of Q-factor variance in a data set. Note that a population model based on an additive combination of R- and Q-factors implies that a row-centered matrix of individual R-factor scores is combined with a row-and-column-centered matrix of individual Q-factor scores (Equations 15, 16). Burt (1937) demonstrated that the eigenvalues of R- and Q-factor analysis of a row-and-column-centered matrix are identical, so that a high similarity of eigenvalues should be expected for combined R- and Q-factor models, even when the resulting matrix is not perfectly column-centered. Therefore, Q-factor analysis will yield a number of substantial eigenvalues, even when the data can perfectly be described by R-factors analysis. Thus, the eigenvalues of Q-factor analysis do not inform unambiguously on the amount of Q-factor variance.

It is therefore proposed to consider the bivariate scatterplot of observed variables in order to ascertain whether between-subject variance that could be due to R-factors is combined with a substantial amount of within-subject variance that could be due to Q-factors. Different within-subject profiles that might be caused by $q_Q > 1$ Q-factors imply that not all differences between two observed z-standardized variables $z_1$ and $z_2$ are equal. For $q_Q = 2$, for example, there could be one group of participants with $z_1 - z_2 > 0$ and a second group with $z_1 - z_2 < 0$. It follows that the variance of the z-score differences $d$, $\sigma_d$, is greater zero for $q_Q \geq 2$. According to Rodgers and Nicewander (1988, p. 64) the correlation can be written as

$$\rho_{z_1 z_2} = 1 - \sigma_d^2 / 2. \tag{17}$$

As $q_Q \geq 2$ implies $\sigma_d > 0$, it follows from Equation 17 that $\rho_{z_1 z_2} < 1$. An example for $n = 145$ cases and $q_Q = 3$ is given for $r_{z_1 z_2} = .40$ in Figure 3 (dots). The concentration of points on three lines is extreme for $q_Q = 3$, so that the bivariate distribution is quite different from the bivariate distribution for the same correlation and $q_Q = 0$ (Figure 3, crosses). For $q_Q = 0$ there is a bivariate normal distribution, which is clearly not the case for $q_Q = 3$. As the distributions in Figure 3 are not skewed, only tests of the multivariate kurtosis were performed with the macro provided by DeCarlo (1997) at $\alpha = .05$. Srivastava's (1984) test for multivariate kurtosis ($\beta_{2,p} = 2.26$, $N(\beta_{2,p})=-2.59, p < .01$), Small's (1980) test of multivariate kurtosis ($Q_2=298.95, df = 2, p < .01$), and Mardia's (1970) test indicate a significant departure from multivariate normal kurtosis ($\beta_{2,p} = 6.36, N(\beta_{2,p})=-2.47, p < .05$). The example shows that a bivariate distribution clearly based on $q_Q = 3$ may result in a platykurtic departure from the kurtosis of the bivariate normal distribution. Even when different reasons for platykurtic multivariate distributions are possible, tests of the multivariate kurtosis might be useful indicators of $q_Q > 1$. Visual inspection of scatterplots may be performed when significant departures from the multivariate normal distribution occur.

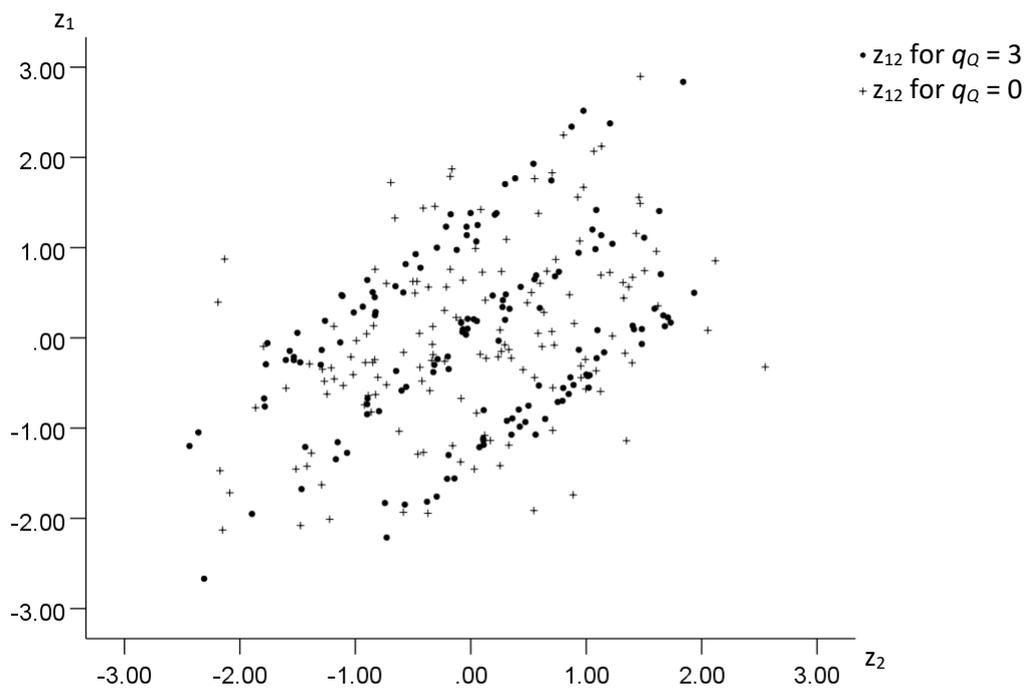

**Figure 3.** Scatterplot of $z_1$ and $z_2$, $n = 145$, $q_Q = 3$, and $r_{z_1.z_2} = .40$.

In order to investigate the usefulness of tests for the kurtosis of the multivariate normal distribution as indicators for $q_Q > 1$, tests were performed for $q_R = q_Q = 3$ and $p = 15$. The tests were based on 2,000 samples with $n = 300, 600$ and $900$ with $\lambda_Q = .90$, $\lambda_R = .50$ and $.70$, $w_R^2 = 0.10, 0.25, 0.50, 1.00$, and $\alpha$-levels of $.05, .10$, and $.20$. As the test is employed in order to evaluate conditions for R-factor analysis, an alpha-level beyond the conventional .05-level might be justified. Note that the $w_R^2 = 1.00$ condition is a condition without any effect of Q-factors, so that no detection rate beyond chance level should be expected for this condition. Overall, the highest detection rates for data with substantial Q-facror variance were found for Mardia's coefficient (see Table 2). However, for $n = 300$ and $w_R^2 = 1.00$ the rate of false positives is slightly above chance for Mardia's coefficient. As the power for the identification of substantial Q-factor variance was sufficiently high for Srivastava's and Small's tests without substantial false positives $w_R^2 = 1.00$, these tests might be recommended.

**Table 2.** Percentage of $p$-values of tests of kurtosis indicating significant departures from multivariate normality at alpha = .05, .10, and .20, for $n = 300$, 600, and 900, $p = 15$, $\lambda_R = .50$ and .70, $\lambda_Q = .90$

| $w_R^2$ | $\lambda_R$ | Method | $n = 300$ | $n = 600$ | $n = 900$ |
|---|---|---|---|---|---|
| | | | $\alpha = .05 / .10 / .20$ | $\alpha = .05 / .10 / .20$ | $\alpha = .05 / .10 / .20$ |
| | | Small | 76.7 / 83.6 / 90.6 | 97.7 / 99.1 / 99.5 | 99.9 / 99.9 / 99.9 |
| 0.25 | .50 | Srivastava | 84.3 / 87.5 / 90.5 | 99.6 / 99.8 / 99.9 | 100.0/100.0/100.0 |
| | | Mardia | 97.6 / 98.4 / 99.0 | 100.0/100.0/100.0 | 100.0/100.0/100.0 |
| | | Small | 35.1 / 43.9 / 56.2 | 66.9 / 76.1 / 84.2 | 84.8 / 89.7 / 93.6 |
| 0.50 | .50 | Srivastava | 53.6 / 61.3 / 69.9 | 90.1 / 93.1 / 95.6 | 98.7 / 99.2 / 99.6 |
| | | Mardia | 79.3 / 83.3 / 88.0 | 99.8 / 99.9 / 99.9 | 100.0/100.0/100.0 |
| | | Small | 8.1 / 15.0 / 26.2 | 13.3 / 21.0 / 33.6 | 18.0 / 26.4 / 39.8 |
| 0.75 | .50 | Srivastava | 10.7 / 17.4 / 28.1 | 24.3 / 31.7 / 43.2 | 39.3 / 47.2 / 58.8 |
| | | Mardia | 15.7 / 23.1 / 33.7 | 51.3 / 59.6 / 69.4 | 76.0 / 81.6 / 87.7 |
| | | Small | 6.2 / 11.3 / 21.0 | 5.3 / 10.8 / 20.9 | 5.8 / 9.9 / 19.8 |
| 1.00 | .50 | Srivastava | 3.4 / 8.7 / 20.1 | 4.7 / 9.6 / 18.8 | 5.1 / 10.6 / 21.5 |
| | | Mardia | 7.5 / 15.2 / 27.9 | 6.8 / 12.7 / 24.5 | 5.8 / 11.3 / 22.8 |
| | | Small | 34.4 / 43.4 / 56.9 | 65.9 / 73.8 / 82.7 | 83.2 / 88.8 / 93.5 |
| 0.25 | .70 | Srivastava | 85.1 / 88.7 / 92.4 | 99.5 / 99.6 / 99.8 | 100.0/100.0/100.0 |
| | | Mardia | 96.6 / 97.5 / 98.6 | 100.0/100.0/100.0 | 100.0/100.0/100.0 |
| | | Small | 13.3 / 21.1 / 33.7 | 25.0 / 35.5 / 48.5 | 35.8 / 46.2 / 59.4 |
| 0.50 | .70 | Srivastava | 54.8 / 62.7 / 71.0 | 89.8 / 92.6 / 95.8 | 98.6 / 99.3 / 99.5 |
| | | Mardia | 75.2 / 80.2 / 85.3 | 99.3 / 99.7 / 99.9 | 100.0/100.0/100.0 |
| | | Small | 6.5 / 18.1 / 22.8 | 7.9 / 14.5 / 25.2 | 8.4 / 15.6 / 27.6 |
| 0.75 | .70 | Srivastava | 11.9 / 18.1 / 29.0 | 25.1 / 32.4 / 43.5 | 38.4 / 46.4 / 58.0 |
| | | Mardia | 13.8 / 20.9 / 32.2 | 45.0 / 54.4 / 64.4 | 71.2 / 77.9 / 84.2 |
| | | Small | 6.5 / 11.6 / 21.3 | 6.0 / 10.9 / 22.5 | 5.3 / 9.9 / 19.7 |
| 1.00 | .70 | Srivastava | 3.5 / 8.7 / 19.7 | 4.8 / 10.5 / 19.8 | 4.8 / 10.3 / 19.6 |
| | | Mardia | 7.5 / 15.0 / 27.4 | 6.8 / 12.6 / 23.5 | 5.6 / 11.6 / 23.3 |

**Discussion**

As R-factor analysis of variables observed for a large number of individuals is the dominant form of factor analysis in several areas of social sciences, it might happen that R-factor analysis is routinely performed even when the population model comprises R- and Q-factors. For example, in the domain of personality research it has been assumed that Q-factors or type-factors may be relevant in addition to the well-known R-factors (e.g., Gerlach et al., 2018; Gilbert, et al., 2021; Ramos, Mata, & Nacar, 2021). This leads to the question whether performing R-factor analysis of data from a population model comprising R- and Q-factors may result in biased loading estimates. R-factor analysis of data from population models comprising R- and Q-factors were therefore investigated. It was noted that a model comprising R- and Q-factors introduces loading indeterminacy beyond rotational indeterminacy. For such a model, the number of model parameters is substantially larger than the number of elements of the covariance matrix of observed variables. It was shown that R-factor analysis of data based on a population model comprising R- and Q-factors leads to biased R-factor loading estimates. For such data R-factor analysis introduces variability into the loading estimates. Thus, when the observed variables have equal R-factor loadings in a population model comprising R- and Q-factors, the loading estimates resulting from R-factor analysis of the observed variables will typically be unequal. This bias of R-factor loading estimates and the variation of R-factor loading estimates beyond chance level was also shown in a simulation study. It was illustrated in the simulation study that the additional loading variability may hamper factor identification.

As the use of R-factor analysis for data drawn from a population based on R- and Q-factors may result in biased R-factor loading estimates, it might be of interest to detect Q-factor variance in observed variables as a prerequisite of R-factor analysis. As eigenvalues of correlation matrices may be in ambiguous and because Q-factor variance leads to platykurtic multivariate distributions of observed scores, it was proposed to use tests for the multivariate normality as indicators for Q-factor variance. In a simulation study Mardia's test of the multivariate kurtosis was more sensitive for the detection of relevant Q-factor variance than Srivastava's and Small's test. However, a slight tendency of false positive results was also found with Mardia's test so that Srivastava's and Small's test might also be recommended. As different reasons are possible for departures of the kurtosis from the kurtosis of the multivariate normal distribution are possible, an inspection of scatterplots is recommended when a test of the multivariate kurtosis of the data is significant. The inspection of scatterplots may be combined with pairwise tests of the bivariate kurtosis in order to eliminate observed variables with substantial Q-factor variance from R-factor analysis.

To sum up, the present paper is a caveat that population models comprising R- and Q-factors do have loading indeterminacy beyond rotational indeterminacy and that performing R-factor analysis of data based on such models results in biased R-factor loading estimates. The bias is due to the fact that the model of R-factor analysis does not correspond exactly to the population model comprising R- and Q-factors. Tests of the multivariate kurtosis might be used for the detection of Q-factor variance as a prerequisite for R-factor analysis. Further research should compare the effect of model error due to Q-factor variance on the results of R-factor analysis with the effect of model error based on minor factors as it has been discussed by MacCallum (2003). The combined effect of both types of model error based on minor factors and model error based on Q-factor variance might be investigated in future research as it may occur in real data.